\pdfoutput = 1 

\documentclass{article}
\usepackage[T1]{fontenc} 
\usepackage[utf8]{inputenc} 
\usepackage{ismir, amsmath,cite,url}
\usepackage{graphicx}
\usepackage{color}
\usepackage{enumitem}

\usepackage{epsfig}
\usepackage{epstopdf}
\usepackage[bookmarks=false]{hyperref}

\newcommand{\squeeze}{\vspace{-0.1cm}}

\newcommand{\NOTE}{\texttt{NOTE}}
\newcommand{\VEL}{\texttt{VEL}}
\newcommand{\TIME}{\texttt{TIME}}
\newcommand{\SKEL}{\texttt{SKEL}}

\newcommand{\Loss}{\mathcal{L}}

\title{PocketVAE: A two-step model for \\ groove generation and control}

\oneauthor
{Names should be omitted for double-blind reviewing}
{Affiliations should be omitted for double-blind reviewing}
\oneauthor
{Kyungyun Lee, Wonil Kim, Juhan Nam}
{Graduate School of Culture Technology, KAIST \\{\tt \{kyungyun.lee, ianwonilkim, juhan.nam\}@kaist.ac.kr}}

\def\authorname{F. Author, S. Author, and T. Author}

\begin{document}

\maketitle

\begin{abstract}
Creating a good drum track to imitate a skilled performer in digital audio workstations (DAWs) can be a time-consuming process, especially for those unfamiliar with drums. In this work, we introduce PocketVAE, a groove generation system that applies grooves to users' rudimentary MIDI tracks, i.e, templates. Grooves can be either transferred from a reference track, generated randomly or with conditions, such as genres.
Our system, consisting of different modules for each groove component, takes a two-step approach that is analogous to a music creation process. First, the note module updates the user template through addition and deletion of notes; Second, the velocity and microtiming modules add details to this generated note score. 
In order to model the drum notes, we apply a discrete latent representation method via Vector Quantized Variational Autoencoder (VQ-VAE), as drum notes have a discrete property, unlike velocity and microtiming values. We show that our two-step approach and the usage of a discrete encoding space improves the learning of the original data distribution. 
Additionally, we discuss the benefit of incorporating control elements - genre, velocity and microtiming patterns - into the model. 

\end{abstract}

\section{Introduction}\label{sec:introduction}

"Playing in the pocket" is a widely used phrase especially in drum and bass performances, which refers to having a good groove through performance details, such as ghost notes and dynamics. Just as it requires years of experience to play in the pocket, reproducing such nontrivial grooves into a drum track in DAWs also demands a significant amount of time and knowledge about real drum performances as well as the software itself. It is often the case that novice musicians are able to sketch out basic drum patterns, but are struggling to develop them further.

Creating tools to make music creation accessible and to boost creativity is a convincing application for music AI technology. Thus, we aim to develop a groove generation system, which can speed up the process of adding the "pocket" groove to users' skeletal beat ideas. Grooves can be transferred from a reference track or generated by the system. 
In this work, \textit{groove} is represented as a combination of 3 components: decorative notes (e.g. ghost notes), velocity (e.g. accents) and microtiming (e.g. push/pull, laidback)
\cite{greenwald2002hip, fruhauf2013music}.
Though not strictly corresponding to real performance, we consider these components as independent features that can be modified individually. 

For many tasks in audio and music domain, the use of Vector Quantized Variational Autoencoder (VQ-VAE) \cite{oord2017neural} has shown promising results due to the discrete nature of speech and musical scores \cite{baevski2019vq, dhariwal2020jukebox, cifka2021self}. Thus, we apply vector quantization to the latent encoding space to model drum notes, which brings significant performance improvement.
As drum track creation in DAWs can be thought of as a two-step process, we translate this process into the modular design of our system: 1) the \NOTE{} module elaborates on the input template with decorative notes, such as ghost notes and hi-hat subdivisions, and 2) the \VEL{} and \TIME{} module add velocity dynamics and timing to the generated note score.

Our contributions can be summarized as:
\begin{itemize}[leftmargin=*]
\squeeze\squeeze
\item Introducing PocketVAE, a model that learns each groove component with separate modules
\squeeze\squeeze
\item Adopting VQ-VAE for modeling drum notes and showing promising results
\squeeze\squeeze
\item Incorporating conditions (genre, velocity, microtiming patterns) for controllable groove generation
\end{itemize}
\squeeze
Code and samples for this paper are made available\footnote{Code: https://github.com/kyungyunlee/PocketVAE ,\\samples: https://kyungyunlee.github.io/PocketVAE\_demo/}. 



\section{Related works} 

Researches on tools that help beginners express more advanced musical ideas are actively being conducted. However, previous works have largely focused on piano performances. One interesting system is Piano Genie\cite{donahue2019piano}, which allows users to perform improvisation without musical knowledge through only a limited number of keys. 
Other systems enable users to harmonize melodies \cite{huang2019bach} or add performance styles to piano scores \cite{jeong2019virtuosonet, oore2020time, chacon2016basis}. 

Such research is extended to drum performance generation, although there is a comparably limited amount. 
Recently, Gillick et al. \cite{gillick2019learning} presented a series of groove generation systems for applications such as infilling and groove transfer, that elaborate on the reduced beats to generate full performances. They also released a large set of drum performance recordings by professional drummers. Several works explored the modeling of microtiming using various regression methods \cite{wright2006towards} or recurrent neural networks \cite{burloiuadaptive}. Hellmer et al.\cite{hellmer2015quantifying} introduced methods to quantify the measures of microtiming of a drum track, which is also adopted in this work. 
Although not performance-related, there exist other works on drums, such as beat generation \cite{tokui2020can, wei2019generating}, drum sound synthesis\cite{ramires2020neural, nistal2020drumgan} and interface design\cite{vogl2016intelligent, bruford2019groove}. 

%

We draw a parallel between our task and image-to-image translation, in which one form of image is transformed into another. Often using generative models, such as conditional GANs, there have been works on converting images from one condition to another\cite{zhu2017toward}, from night to day for instance, or transforming semantic label maps to photographic images \cite{chen2017photographic, wang2018high}. The latter example particularly resembles our task of creating a more complex output based on a coarser, but structured input.








\squeeze
\squeeze

\section{Methods}

\begin{figure}[t]
\centering
 \includegraphics[width=\columnwidth]{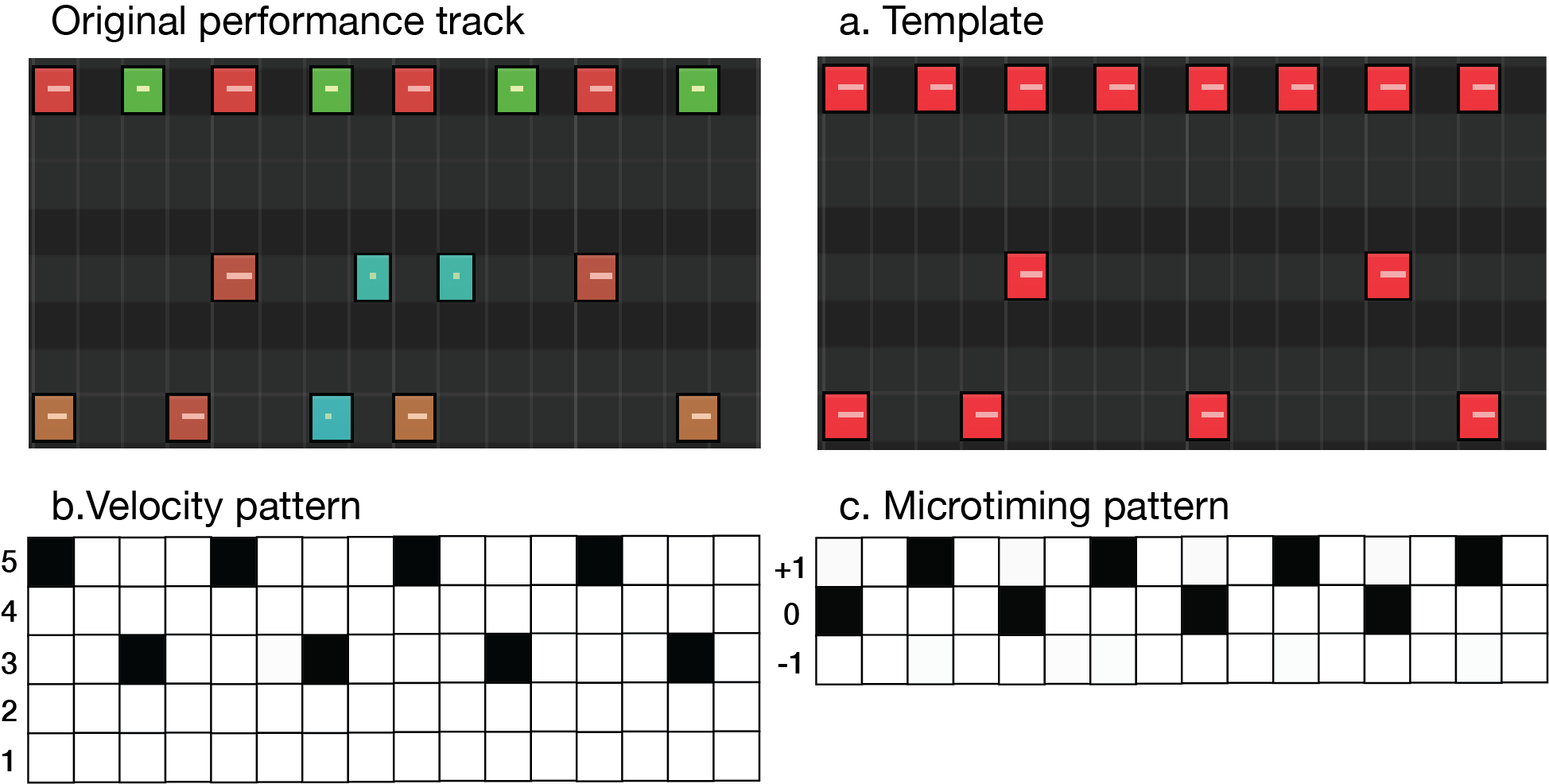}
\caption{Derivation of the template and pattern matrices from original track during the data processing step. Images on the first row is a capture of the drum tracks from the DAW, Logic Pro X. Each row indicates different drum classes (i.e. Hi-hat) and each column indicates timestep of 16th notes. Colors indicate velocity values, where red means high velocity (max = 127) and blue means low velocity. Template (a) is a simplified version of the original performance track, removing low velocity notes and setting velocity values to 127. Velocity pattern (b) indicates the velocity trend of the cymbal notes, where velocity values are bucketed into 5 classes with 5 being the highest velocity class. Microtiming pattern (c) represents the average microtiming value for all notes played in the same timestep. Positive, "+1", indicates that the overall feel is laidback, while negative, "-1", indicates that the feel is pushed. }
  \label{fig:data_processing}
\end{figure}

\subsection {Data collection}
Our dataset consists of 2841 tracks from GrooveMIDI\cite{gillick2019learning}, Groove Monkee\footnote{https://groovemonkee.com/}, BFD3\footnote{https://www.fxpansion.com/products/bfd3/}, and Reddit\footnote{https://www.reddit.com/r/datasets/comments/3akhxy/\\the\_largest\_midi\_collection\_on\_the\_internet/}. 
We collected tracks for the following genres (\# of tracks/\# of 2-bar segments): electronic (451/985), rock (524/4798), funk (288/1064), jazz (155/988), blues (292/2569) and hip hop (1131/8839). 
List of all tracks used in this paper is provided \footnote{https://github.com/kyungyunlee/PocketVAE/tree/master/\\datasets/data\_splits}. 
During collection, we excluded tracks that are not in 4/4 meter and that contain "fill-in" in their file names as they refer to files with only drum fills. Also, we did not use tracks with less than 3 instruments and with no kicks or snares, unless their genre is jazz, since typical drum patterns are composed of at least kick, snare and cymbal. 

\squeeze

\subsection{Data pre-processing}

Data pre-processing involves 1) converting MIDI files into model-friendly data matrices, 2) deriving templates and 3) obtaining velocity and microtiming patterns. We follow the General MIDI percussion map and reduce the number of drum instruments to 7 major classes: kick, snare, hi-hat closed, hi-hat open, ride, crash and tom. Although 16th note resolution cannot represent shorter notes, such as snare rolls, we find 16th note resolution sufficient for our data representation. All tracks are segmented into 2-bar sequences with 1-bar sliding window over the full track. Shorter tracks are repeated to meet the minimum 2-bar requirement. We chose 2 bars, as it is enough length to explore grooves in electronic music production. 

\noindent\textbf{MIDI $\mapsto$ <Note, Velocity, Microtiming>}
Our MIDI-to-matrix format adheres to that of the previous work\cite{gillick2019learning}. A MIDI file is decomposed into 3 matrices: note ($N$), velocity ($V$), microtiming ($M$). Each matrix is a 2-D data $\in R^{T \times I}$, where $T=32$ and $I=7$. Therefore, value at index $[t, i]$ refers to the 16th note position $t$ and the instrument index $i$. Matrix $N$ is binary-valued, where 1 indicates the presence of a drum hit. Values of $V$ are real numbers in [0, 1], which is mapped from the original integer velocity in [0, 127]. In case of $M$, values are also in real numbers, indicating the amount of onset deviation before or after the beat position. Its values are converted from (-60, 60] ticks to (-1, 1] (16th note = 120 ticks).

\noindent\textbf{Note $\mapsto$ Template} 
The purpose of deriving templates is to mimic user inputs. We apply reduction per drum instruments for each track in the dataset to obtain a template matrix, $P \in R^{T \times I}$ (Figure \ref{fig:data_processing}). We aim to keep only the salient notes that represent the underlying rhythm of the original track. For cymbals (hi-hats, ride and crash), we first find the dominant cymbal and determine whether the track's rhythm is in 8th or 16th beat. 
We count the number of cymbal hits and if it is greater than 3/4 of the sequence length, the track is assumed to be in the 16th beat rhythm. Otherwise, it is in the 8th beat. Depending on the result, either every 16th or 8th position of the main cymbal notes is set to 1. 
In terms of snares, we remove the ghost notes as much as possible, since we consider them as part of the groove expression. We compute the average snare velocity on the second and fourth quarter-note positions ("two four") and use it as the threshold for removing ghost notes. For kicks, we consider them to contribute largely to the main rhythm, thus we try to keep as much notes as possible. We arbitrarily set the threshold value to 40 (max velocity is 127) and keep the notes above this value. Lastly, we do not keep any notes on the tom.  

\noindent\textbf{Velocity, Microtiming $\mapsto$ Velocity, microtiming patterns}
We derive velocity and microtiming patterns, $C_V \in R^{T \times 5}$ and $C_M \in R^{T \times 3}$, from $V$ and $M$ matrices. These pattern matrices (Figure \ref{fig:data_processing}) are like drawing contours of the velocity and microtiming patterns for an intuitive manipulation of dynamics. Velocity pattern is only obtained from the cymbal velocity, and values are quantized into 5 classes, [0.2, 0.4, 0.6, 0.8, 1.0].
Microtiming values are an average of all values at each time step, mapped to 3 classes, [-1, 0, 1], where -1 means "pushed", 0 means played on the actual beat position and +1 means "laidback." 
We observe improvements in performance with these extra guidance for the model .




\subsection {Baseline models}
\subsubsection{K-NN}
We implement a K-nearest neighbor (K-NN) model, similar to the previous literature \cite{gillick2019learning}. Given an input template, K-NN model searches for tracks with the most similar templates in the dataset and returns the aggregation of the top K tracks. K-NN is implemented with Faiss\footnote{https://github.com/facebookresearch/faiss} and inner product is used as the similarity metric. K is set to 20. 
The resulting note matrix $N_{out}$ is computed by taking the majority vote at each index. 
\begin{equation}  \label{eq:cosine}
N_{out} = 
\begin{cases}
    1, & \text{if } \sum_{k=0}^{K-1} N_k \geq \frac{K}{2} \\
    0,              & \text{otherwise}
\end{cases}
\end{equation} 
Velocity and microtiming matrices, $V_{out}$ and $M_{out}$, are each calculated by averaging that of its K neighbors. 
\begin{equation}  \label{eq:cosine}
V_{out} = \frac{1}{K} \sum_{k=0}^{K-1} V_k
\end{equation} 

\squeeze
\squeeze

\subsubsection{One-step model}
The "one-step" model, is a reduced version of our two-step model (section \ref{sec:two-step}). It consists of a single variational autoencoder (VAE), which learns to generate three groove components all at once. This is similar to the Seq2Seq model from \cite{gillick2019learning}. The model is conditioned on template matrices to learn to disentangle the groove from the input template. However, the limitation of this model is prominent when we want to generate new groove, because a single latent vector cannot disentangle each groove component, thus lacking controllability. 

\squeeze
\squeeze 

\subsection{Proposed Two-step Model}\label{sec:two-step}

\begin{figure}[t]
\centering
 \includegraphics[width=\columnwidth]{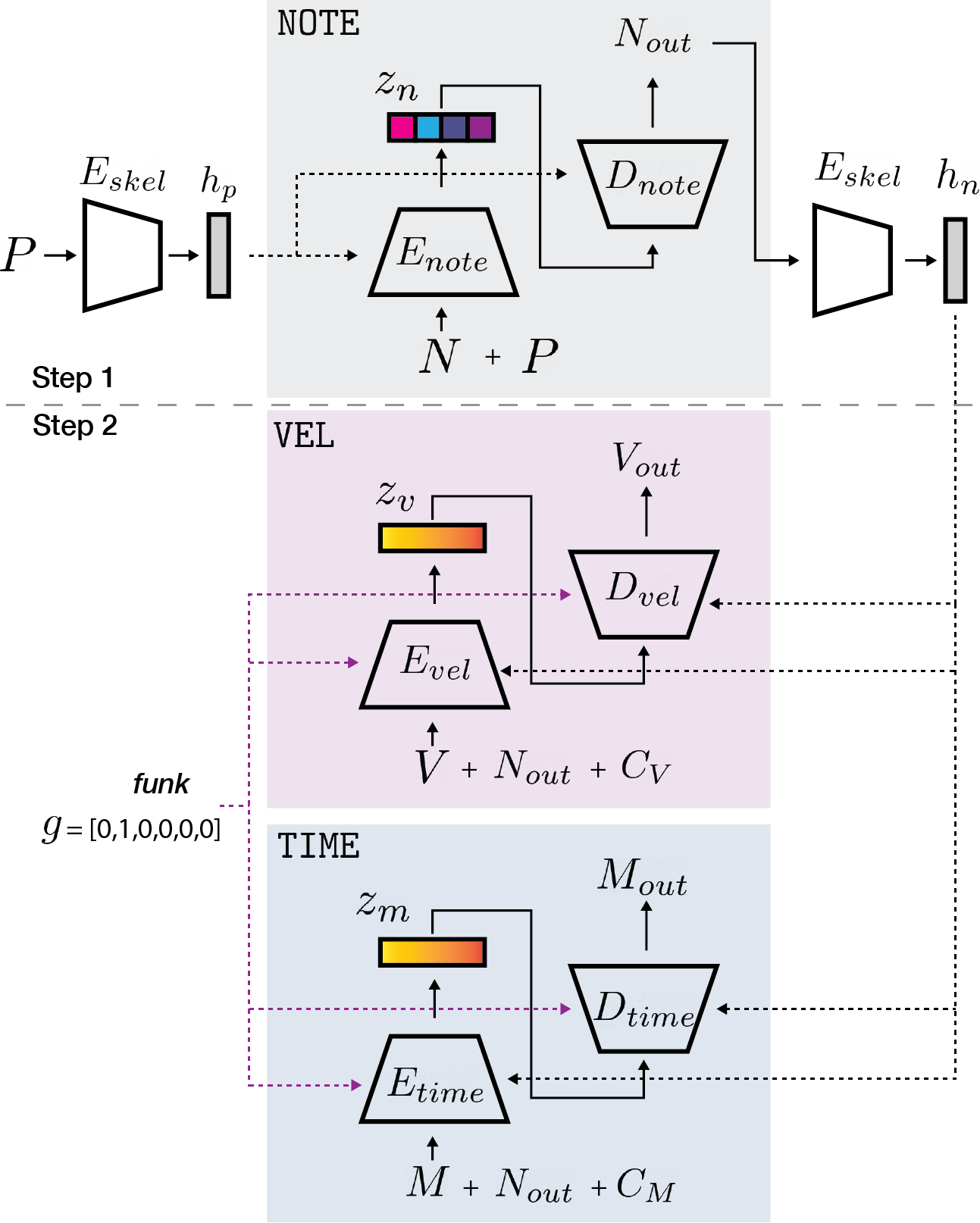}
\caption{The Two-step PocketVAE architecture with genre condition (ex. funk). Short dashed lines refer to conditional inputs to GRUs as hidden vectors.}
  \label{fig:model}
\end{figure}

The proposed two-step model, also called "PocketVAE", consists of 3 main modules for each groove component - \NOTE{}, \VEL{} and \TIME{} - and a skeleton module, \SKEL{} (Figure \ref{fig:model}). 
We also introduce variations of this model by adding control elements as conditions with genre and/or velocity and microtiming patterns. 
The model generates in a cascaded manner, in which the output of the \NOTE{}, $N_{out}$, is used as a conditional input to the \VEL{} and \TIME{} modules, $\VEL(V|N_{out})$ and $ \TIME(M|N_{out})$; hence, the name "two-step". 
The \NOTE{} module itself, $\NOTE(N|P)$, is conditioned on the template, $P$. Conditioning with $P$ and $N_{out}$ are done by concatenating with the input matrices, as well as initializing the hidden layers with the encoded vectors from the \SKEL{} encoder, $h_n$ and $h_p$. 
These conditions guide the network to extract only the relevant information that contributes to the groove. 
Modeling each component with separate modules naturally brings disentanglement in the groove feature space, allowing us to trivially add control elements per component via velocity and microtiming patterns (section \ref{section:model_control}).

\subsubsection{The \SKEL{} sub-module}
The \SKEL{} sub-module is an encoder that compresses the template, $P$, or the note matrix, $N$, into a single latent vector, $h_n, h_p \in R^{16}$. The sub-module consists of a single bidirectional gated recurrent unit (GRU) \cite{chung2014empirical} layer with 128 hidden units. The encoded template, $h_p$, is used to initialize the hidden layers of the \NOTE{} module, while the encoded note matrix, $h_n$, is used to initialize the hidden layers of \VEL{} and \TIME{} modules (Figure \ref{fig:model}). 

\subsubsection{The \NOTE{} module}
The \NOTE{} module learns to manipulate the notes of the input template through addition and deletion of notes. It is a conditional VQ-VAE that encodes the binary note matrix, $N$, into a series of latent variables $\boldsymbol{z}_{N} = c_1,...,c_S$, where $ c_s \in R^D$ and $S=8$. Encoded variables are discretized by the codebook $H(z_{N}) \mapsto e_{N}$, in which $H \in R^{K \times D}$ with $K=64$ codes each with $D=16$ dimensions. Then, the decoder reconstructs $N$ from $e_{N}$. 
The encoder, $E_{note}(N|P, h_p)$, consists of a bidirectional GRU with 256 hidden units followed by 3 1-D convolutional layers (architecture details can be seen here\footnote{https://github.com/kyungyunlee/PocketVAE/tree/master/\\src/models/README.md}).
Convolution operation reduces the dimensions in the time domain and we purposefully choose the kernel size as 4 to summarize 4 16th notes. 
We set the length of code sequence, $S$, to 8, which can be be interpreted as each code representing 4 ($T/S=4$) 16th notes, a reasonable unit for drum loops.  
The decoder, $D_{note}(z_N|P,h_p)$, consists of operations that reverse that of the encoder.

Loss function related to the \NOTE{} module consists of a note  reconstruction loss, a codebook loss and a commitment loss. 
\begin{equation}
\scalebox{1.0}{$\begin{aligned}
    & \Loss_{note} = \Loss{}_{recon} + \beta_1 \cdot \Loss{}_{cb} + \beta_2 \cdot \Loss{}_{cmt} \\ 
    & \Loss_{recon} = BCE(N_{out}, N) \\
    & \Loss_{cb} = \lVert {sg[H(z_{N})] - z_{N}} \lVert _2 ^2\\ 
    & \Loss_{cmt} = \lVert {H(z_{N}) - sg[z_{N}]} \lVert _2 ^2
\end{aligned}$}
\end{equation}
Here, $sg[\cdot]$ refers to stop gradient operator, which prevents gradients from being computed. We set both $\beta_1$ and $\beta_2$ to 0.2. 


\subsubsection{The \VEL{} and \TIME{} modules} 
The \VEL{} and \TIME{} modules are expected to learn the velocity dynamics and timings of the original track. They are conditional VAEs. Encoders of each module, $E_{vel}(V|N, h_n)$ and $E_{time}(M|N, h_n)$, are composed of a bidirectional GRU with 256 hidden units, each outputting a single 64-dimensional vector, $z_{V}, z_{M} \in R^{64}$. Decoders, $D_{vel}(z_{V}|N, h_n)$ and $D_{time}(z_{M}|N, h_n)$, are 2 layers of unidirectional GRU with 256 hidden units, trained to predict the values at the next time step. In the last layer, Sigmoid function is applied for velocity predictions and Tanh function for microtiming predictions. 
Loss functions for both models are composed of a mean squared error (MSE) and a regularization term, Kullback-Leibler (KL) divergence.

\squeeze
\begin{equation}
\begin{aligned}
     \Loss_{vel} =  \lVert {V - V_{out}} \lVert _2^2 + \beta \cdot KL\\ 
     \Loss_{time} = \lVert {M - M_{out}} \lVert _2^2 + \beta \cdot KL
\end{aligned}
\end{equation}

Putting all modules together, we train them jointly with the following combined loss function. 

\squeeze

\begin{equation}
    \Loss{} = \Loss{}_{note} + \Loss{}_{vel} + \Loss{}_{time} 
\end{equation}

During training, we exponentially decrease the teacher forcing ratio from 1 to 0.5 and increase the $\beta$ value for the regularization term from 0 to 0.2. We use Adam as our optimizer. 
To evaluate the effects of using VQ-VAE, we also implement a model with the \NOTE{} module as a regular VAE (Referred to as 2-step (VAE)).

\subsubsection{Learning the prior}\label{sec:prior}
Learning the prior distribution of code sequences is required for generating new groove. After training the three modules, we train the prior model after training the PocketVAE, to learn the prior distribution of the codes. The model consists of a 2-layer GRU with genre condition and is trained autoregressively to predict next code. We use cross entropy as our loss function. As a result, during inference, a new code sequence can be sampled given a genre. 

\subsubsection{Adding control elements} \label{section:model_control}
Genre is represented as an one-hot vector, $g \in R^6$, and is used to initialize the hidden layers of encoders and decoders in the model. For the \NOTE{} module, we incorporate genre information only in the prior module.  
Velocity and microtiming patterns, $C_V$ and $C_M$, are concatenated with the inputs of the encoder and decoder of \VEL{} and \TIME{} modules, respectively. Therefore, conditioning the \VEL{} module with genre and velocity pattern is represented as $E_{vel}(V|N,C_V,g)$ and $D_{vel}(z_V|N, C_V,g)$.
\squeeze

\section{Evaluation and Results}
\begin{figure}[t]
\centering
 \includegraphics[width=0.8\columnwidth]{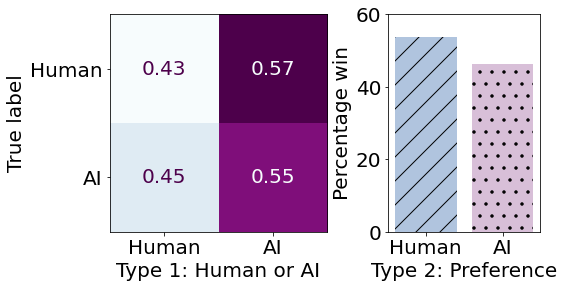}
\caption{Listening test results on two types of questions.}
  \label{fig:listening_test}
\end{figure}

\begin{table*}[t]
\centering
\small
\setlength\tabcolsep{4.5pt}
\begin{tabular}{l|c|c|c|c|c|c}
          & Note F1  & Vel. KL & Vel. MSE & MT KL & MT MSE & Genre Acc. \\ \hline\hline
Test data & - & - & - & - & - & $0.906 \pm 0.042$ \\ 
KNN &  $0.561\pm0.029$ & $0.217\pm0.025$ &  $0.037\pm0.003$ & $0.191\pm0.040$ & $0.011\pm0.001$  & $0.487\pm0.018$  \\

1-step &  $0.816 \pm 0.015$ & $0.115 \pm 0.017$ & $0.025 \pm 0.001$ & $0.846 \pm 0.072$ & $0.013 \pm 0.001$  & $0.554 \pm 0.033$ \\
\hline
2-step (VQ-VAE) & $\mathbf{0.973 \pm 0.004}$ &  $\mathbf{0.004 \pm 0.003}$ & $0.002 \pm 0.000$ & $0.0416 \pm 0.016$ & $0.003 \pm 0.000$ &  $\mathbf{0.866 \pm 0.028}$ \\

+genres & $\mathbf{0.973 \pm 0.000}$ & $\mathbf{0.004 \pm 0.003}$ & $0.002 \pm 0.000$ & $0.0439 \pm 0.016$ & $0.003 \pm 0.000$ & $\mathbf{0.898 \pm 0.029}$ \\

+genres+patterns & $0.972 \pm 0.005$ &  $0.011 \pm 0.009$ & $\mathbf{0.001 \pm 0.000}$ & $\mathbf{0.0154 \pm 0.006}$ & $\mathbf{0.002 \pm 0.000}$ & $\mathbf{0.899 \pm 0.037}$ \\ 

2-step (VAE) & $0.840 \pm 0.024$ & $0.3664 \pm 0.096$ & $0.0020 \pm 0.000$ & $0.0704 \pm 0.027$ & $0.004 \pm 0.000$ & $0.807 \pm 0.026$ \\

\end{tabular}
\caption{Reconstruction metrics are cross validated with 5 random train/valid/test splits. Test data is from the dataset and therefore, is an upper bound.} 
\label{table:model_accuracy}
\end{table*}

\begin{figure*}[t]
\centering
 \includegraphics[width=\textwidth]{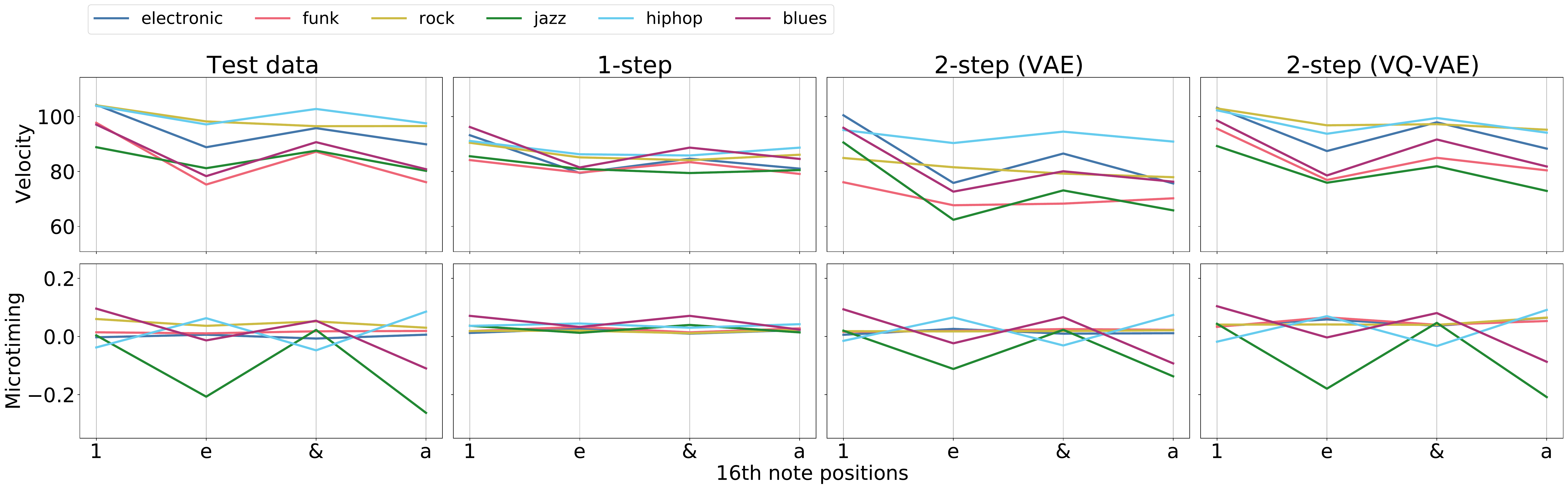}
\caption{Average velocity and microtiming values of the reconstructed test data at each metrical position. "Test data" refers to the original test data.}
  \label{fig:metrical}
\end{figure*}

\subsection{Listening tests}
A total of 24 responses were collected, where 75\% of the participants were either music majors or aficionados. 45.8\% of the participants have played drums before and also, 58.3\% of them have some experience with creating music with DAWs. 16.7\% were professional drummers and/or musicians with more than 7 years of experience in music production. 
We asked participants to answer 2 types of questions. The first type, "Human or AI", asked whether the given track is played by humans or AI. This is to check whether AI generated results sound as natural as human's grooves. The second type, "Preference", gives a template along with grooves generated for the template by both AI and human. Participants were asked to choose the groove they prefer. This simulates the real-life user experience if the model were to be used as a plug-in. Each type consisted of 10 questions. 

For "Human or AI", 45\% of AI performances were incorrectly predicted as human's, but in fact, 57\% of the human performances were also mistaken as AI's (Type 1 in Figure \ref{fig:listening_test}). Even from participants comments, we observed that there were substantially high confusion between tracks by humans and AI. In case of "Preference" questions, a slightly more votes were given to the grooves done by AI than humans (53.7 vs 46.3\%), which is an impressive outcome (Type 2 in Figure  \ref{fig:listening_test}).   
Participants left several interesting comments. One participant reviewed that "Instead of all being mediocre, some were really good and some were quite bad." Some set their own standards to distinguish AI, for instance, "I assumed grooves with little slips were done by AI" and "For some reason, I felt like humans had more context in their playing... beats that were unclear and confused were thought to be AI. Also, I considered very stylish offbeats as done by humans, while ambiguous ones done by AI." There were comments about the desire to use our model if developed as a plug-in - "As a non-drummer, it's hard for me to express the minute details of velocity dynamics, so [the model] would be helpful as a VST." 

\squeeze
\squeeze 

\subsection{Quantitative evaluation}
As evaluation of music generation system often does not directly translate into the measurement of human perception, we try to evaluate the performances in various aspects. Test tracks were not seen during training stages.

\noindent\textbf{Reconstruction}
With a series of metrics, we evaluate each model's ability to reconstruct the original data, which also implies the quality of information it can encode in its latent variables (Table \ref{table:model_accuracy}). Note F1-score is the average F1-score computed at each timestep. It is analogous to that of multi-label classification. Velocity and  microtiming predictions are analysed with two different metrics: KL distance and MSE. KL distance compares the distributions of values between the original and the reconstructed data \cite{gillick2019learning, hellmer2015quantifying, tidemann2007imitating}. Drum performance emphasizes dynamics at each metrical position - 16th note position in a single quarter note beat, referred to as "1, e, \&, a". For instance, accents on the "1" and "\&" will convey a different feel compared to putting them on "e" and "a". The velocity and microtiming values from the entire test data is collected and compared per metrical position. We show the average of that from all four positions in Table \ref{table:model_accuracy}. We provide a visualization in Figure \ref{fig:metrical},  which plots the trends of average velocity and microtiming values at each metrical position per genre. MSE is also computed between the original and the reconstructed values. Lastly, we pre-train a genre classifier, a 2-layer bidirectional GRU, that uses $N, V, M$ matrices to classify the track into one of the six aforementioned genres. The accuracy on the original test data is the upper bound (first row in Table \ref{table:model_accuracy}). 

As shown in Table \ref{table:model_accuracy}, the most significant performance improvement comes from the implementation of VQ-VAE for the \NOTE{} module. Around 30\% increase in the note prediction score comes from using VQ-VAE to learn grooves of drum note patterns. This suggests that modeling drum note information, which are represented as a binary-valued matrix, as a single Gaussian distribution poses a significant limitation. Thus, allowing the use of a multi-modal distribution through the discretization of the latent encoding space shows more promising results. Also, a general trend of performance improvement can be observed when 1) upgrading the one-step VAE to the two-step PocketVAE and 2) adding controls, like genres. The two-step approach especially brings improvement in velocity and microtiming MSEs and microtiming KL distance, as each model can focus on each component, instead of a single model learning them all together. 

Adding conditions to the model not only brings control, but also notable improvement in KL distance particularly for both velocity and microtiming values. We believe that KL distance reflect the global trend of dynamics better than MSE, which can also be observed by drawing metrical trends as in Figure \ref{fig:metrical}. As genres innately exhibit unique trends, performance improvement that comes with the addition of genre condition is sensible. 

We bring attention to the fact that KL distance and MSE metrics do not show correlation, which suggests that a single metric can lead to an incomplete evaluation of the model performance.






\begin{figure}[t]
\centering
 \includegraphics[width=\columnwidth]{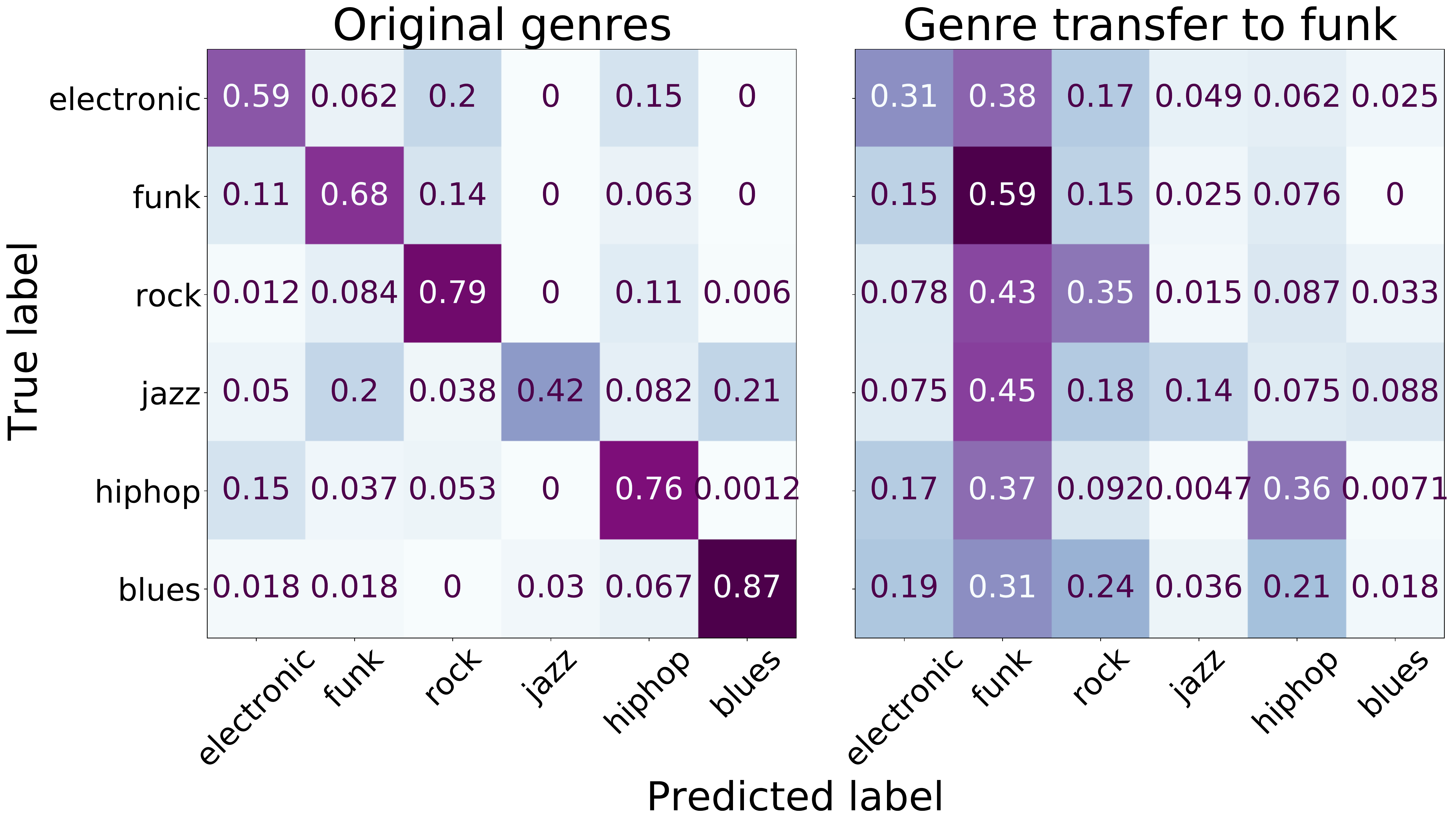}
\caption{Left: Genre classification result on generated data from genres and skeleton in the test data. Right: Same, but on generated outcomes after conditioning all genres as funk. Two-step(VQ-VAE)+genres model is used.}
  \label{fig:genre_transfer}
\end{figure}

\begin{figure}[t]
\centering
 \includegraphics[width=0.9 \columnwidth]{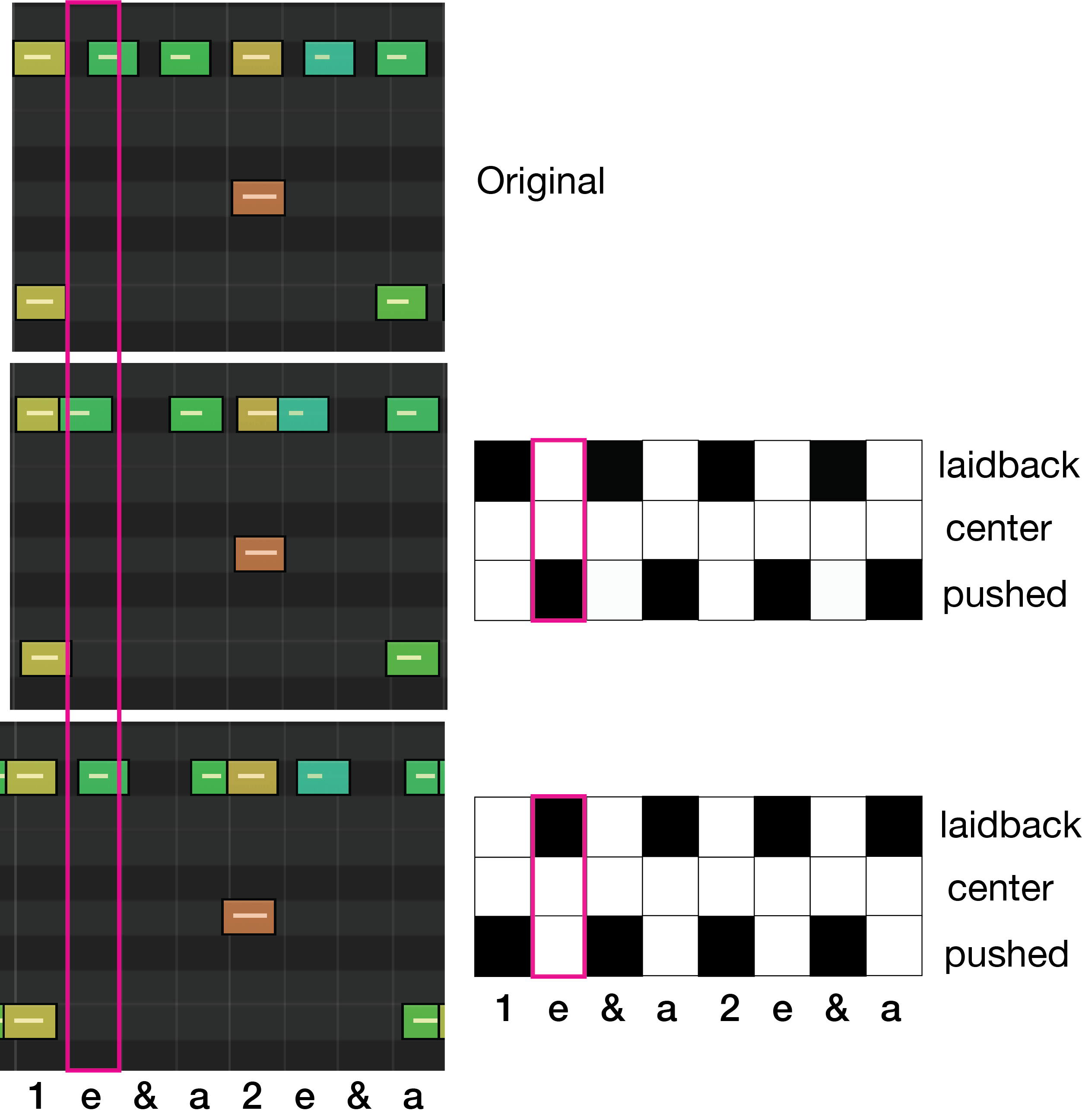}
\caption{Changes in microtiming values as the conditioning patterns are changed via the control template shown on the right-side figures. The control template is a binary valued matrix, where black indicates value 1. We inserted red squares to show the quantized beat location. Now, looking at the second timestep (second column) of the top-most figure, the original performance plays in a laidback style. However, when the control template is changed to "pushed", the model generates a note that is played slightly earlier than the quantized beat (Figure on the second row). Additionally, when the control is changed to "laidback", the model indeed generates the note in a laidback style, which is slightly delayed from the quantized beat. }

  \label{fig:time_pattern}
\end{figure}

\begin{figure}[t]
\centering
 \includegraphics[width=\columnwidth]{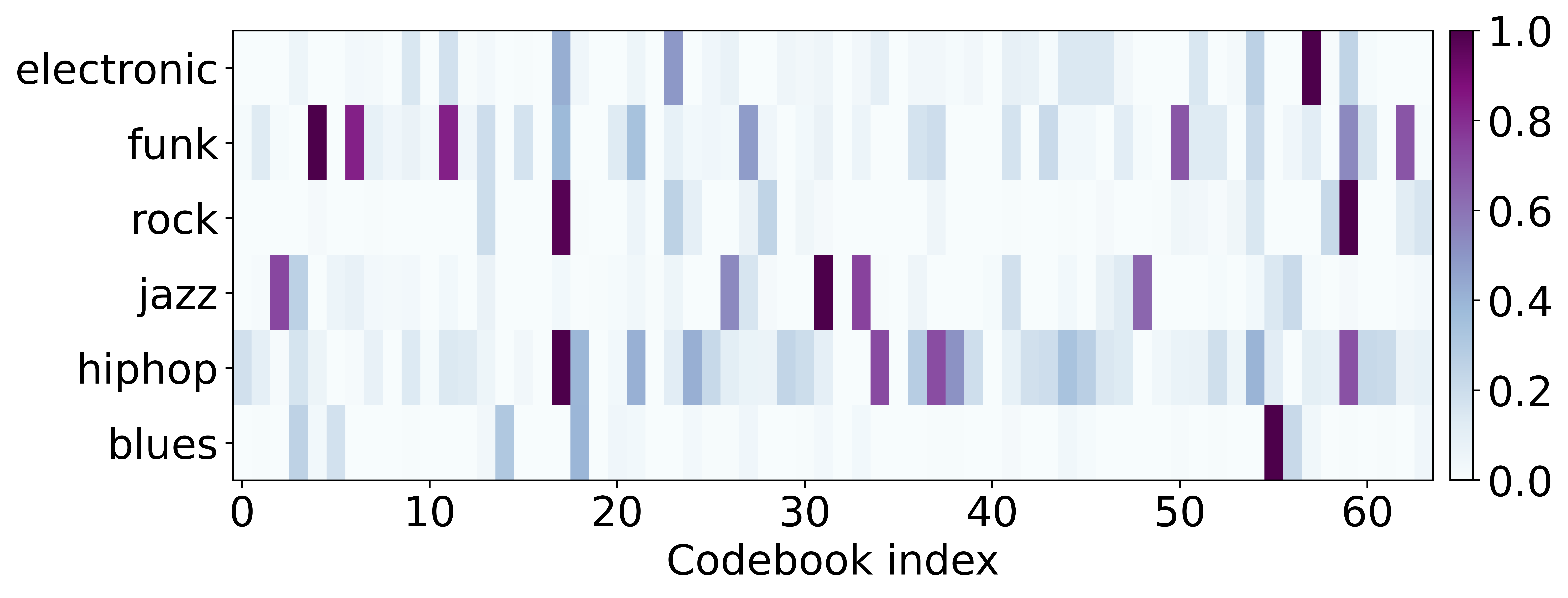}
\caption{The distribution of the number of codebooks that appear per genres when generating code sequences using the prior model (section \ref{sec:prior}). Values are normalized per genres. This shows that certain codes represent one genre more than another. For instance, for funk, code number 4, 6, 11, 50, 59 and 62 appear often.}
  \label{fig:codebook}
\end{figure}

\noindent\textbf{Controllability}
When trained with genre condition, the model should generate grooves in the style of the input genre. Analysis for genre control is done by checking whether altering the input genre will lead to new style of grooves. 
We chose to condition on \textit{funk} and generate groove from templates in the test data.
Then, we performed genre classification and compared the result with that of original-genre-conditioned tracks. 
In the confusion matrix shown in Figure \ref{fig:genre_transfer}, we see that with the \textit{funk} condition, the majority of the tracks are re-classified as \textit{funk}. 

In the previous "Reconstruction" section, we mentioned the improved reconstruction performance in terms of KL distance with use of velocity and microtiming patterns. Here, we qualitatively observe and listen whether providing different patterns modifies the generated groove. In Figure \ref{fig:time_pattern}, we show the generated tracks when conditioned with two different microtiming patterns. Indeed, the highlighted red box shows that with the "pushed" microtiming condition, the note is played slightly before the beat, while with the "laidback" condition, the note is played behind the beat.

Although we explore these types of control, we pose a question on how much control is "just the right amount" for users. For instance, novice users may be expecting the model to fully generate groove from scratch without them having to control anything, while more advanced users may want to incorporate their musical knowledge and customize the model. Too much control means users are expected to have substantial domain knowledge, which could defy the purpose of using the AI-based system. Therefore, the degree of control should be designed with specific target user in mind. In this work, we focused on novice users and have only explored a limited variety of control, but we hope our approach can lead to new ideas in developing various controllable music generation systems.  

\noindent\textbf{Codebook interpretation}
Although it is not easy to interpret what information is exactly contained in each code, we speculate that a single code roughly contains a summary of 1 beat (1 quarter note). We provide examples of generated tracks using each code in our demo page\footnote{https://kyungyunlee.github.io/PocketVAE\_demo/}. Also, we observe different codebook distributions per genres (Figure \ref{fig:codebook}. We find interpreting discrete latent codes more intuitive than continuous latent variables.

\squeeze
\squeeze
\squeeze

\section{Conclusion} 
We introduced PocketVAE, a two-step groove generation and control model. Through learning each groove component - note, velocity and microtiming - individually and discretizing the latent space for note representation, we were able to improve the quality of both reconstruction and overall groove patterns. Although we were able to see better results, there remains a room for improvement. For instance, we found the model lacking in replicating the detailed variance of velocity and microtiming within a track, which may be related to the choice in the loss function.
As our future work, we plan to examine user interface design and plug-in development, which includes models for drum transcription\cite{callender2020improving} and drum sound search\cite{kim2020drum}. As users often find reference tracks in audio, not as MIDI files, we envision such complete groove generation system can one day be embedded in commercial DAWs. 

\section{Acknowledgements}
We thank Jon Gillick for kind comments and reviews.


\bibliography{ISMIR2021_template}

%
%
%
%
%

\end{document}